# Similarity Analysis and Modeling in Mobile Societies: The Missing Link

Gautam S. Thakur*, Ahmed Helmy*, Wei-Jen Hsu+
*Computer and Information Science and Engineering Department, University of Florida, Gainesville, FL
+Cisco Systems Inc., San Jose, CA
{gsthakur,helmy}@cise.ufl.edu, wehsu@cisco.com

## ABSTRACT

A new generation of "behavior-aware" delay tolerant networks is emerging in what may define future mobile social networks. With the introduction of novel behavior-aware protocols, services and architectures, there is a pressing need to understand and realistically model mobile users behavioral characteristics, their similarity and clustering. Such models are essential for the analysis, performance evaluation, and simulation of future DTNs. This paper addresses issues related to mobile user similarity, its definition, analysis and modeling. To define similarity, we adopt a behavioral-profile based on users location preferences using their on-line association matrix and its SVD, then calculate the behavioral distance to capture user similarity. This measures the difference of the major spatio-temporal behavioral trends and can be used to cluster users into similarity groups or communities.

We then analyze and contrast similarity distributions of mobile user populations in two settings: (i) based on real measurements from four major campuses with over ten thousand users for a month, and (ii) based on existing mobility models, including random direction and time-varying community models.

Our results show a rich set of similar communities in real mobile societies with distinct behavioral clusters of users. This is true for all the traces studied, with the trend being consistent over time. Surprisingly, however, we find that the existing mobility models do not explicitly capture similarity and result in homogeneous users that are all similar to each other. Thus the richness and diversity of user behavioral patterns is not captured to any degree in the existing models. These findings strongly suggest that similarity should be explicitly captured in future mobility models, which motivates the need to re-visit mobility modeling to incorporate accurate behavioral models in the future.

## 1. Introduction

Future social networks are expected to have classes of applications that are aware of mobile users' behavioral profiles and preferences and are likely to support peer-to-peer mobile networking including delay tolerant networks (DTNs). A new generation of protocols is emerging, including behavior-aware communication paradigms (such as profile-cast [15]) and service architectures (such as participatory sensing [35, 43]).

Such behavior-aware communication paradigm leverages user behavior and preferences to achieve efficient operation in DTNs (e.g., interest-based target message forwarding; encounter-based routing, mobile resource discovery). Accurate models of mobile user behavioral profiles are essential for the analysis, performance evaluation, and simulation of such networking protocols.

Hence, there is a compelling need to understand and realistically model mobile users behavioral profiles, similarity and clustering of user groups.

Earlier work on mobility modeling presented advances in random mobility models (e.g., RWP, RD [5]), synthetic models that attempt to capture spatial correlation between nodes (e.g., group models [2]) or temporal correlation and geographic restrictions (e.g., Freeway, Manhattan, Pathway Models [1]). More recent models tend to be trace-driven and some account for location preferences and temporal repetition [19]. However, similarity characteristics between clusters of nodes, which lie in the heart of behavior-aware networking, have not been modeled explicitly by these mobility models. Hence, it is unclear whether (and to which degree) similarity is captured.

In this paper, we address issues related to mobile user similarity, its definition, analysis and modeling. Similarity, in this study, is defined by mobility preferences, and is meant to reflect the users interests to the extent that can be captured by wireless measurements of on-line usage. To define similarity, we adopt a behavioral-profile based on users mobility and location preferences using an on-line association matrix representation, then use the cosine product of their weighted Eigen-behaviors to capture similarity between users. This quantitatively compares the major spatio-temporal behavioral trends between mobile network users, and can be used for clustering users into similarity groups or communities. Note that this may not reflect social ties between users or relationships per se, but does reflect mobility-related behavior that will affect connectivity and network topology dynamics in a DTN setting.

We analyze similarity distributions of mobile user populations in two settings. The first analysis aims to establish deep understanding of realistic similarity distributions in such mobile societies. It is based on real measurements of over 8860 users for a month in four major university campuses, USC [38] MIT, Dartmouth [27] and UF. It may be reasonable to expect some clustering of users that belong to similar affiliations, but quantification of such clustering and its stability over time is necessary for developing accurate similarity models. Furthermore, on-line behavior that reflects distribution of active wireless devices may not necessarily reflect work or study affiliations or social clustering. For DTNs, on-line activity and mobility preferences translate into encounters that are used for opportunistic message forwarding, and this is the focus of our study rather than social relations per se.

The second similarity analysis we conduct aims to investigate whether existing mobility models provide a reasonable approximation of realistic similarity distributions found in the campus traces. It is based on existing mobility models where we analyze a commonly used random-based mobility model (the random direction (RD) model [5]) and an advanced trace-driven mobility model; (the time-variant community (TVC) model [19]).

Our results show that among mobile users, we can discover distinct clusters of users that are similar to each other, while dissimilar to other clusters. This is true for all campuses, with the trend being consistent and stable over time. We find an average modularity of 0.64, clustering coefficient of 0.86 and path length of 0.24 among discovered clusters. Surprisingly, however, we find that the existing mobility models do not explicitly capture similarity and result in homogeneous users that are all similar to each other (in one big cluster). This finding generalizes to all other mobility models that produce homogeneous users, not only the mobility models studied in this paper. Thus the richness and diversity of user behavioral patterns is not captured in any degree in the existing models.

Our findings strongly suggest that unless similarity is explicitly captured in mobility models, the resulting behavioral patterns are likely to deviate dramatically from reality, sometimes totally missing the richness in the similarity distribution found in the traces. Furthermore, this indicates our current inability to accurately simulate and evaluate similarity-based protocols, services and architectures using mobility models. This motivates the need to re-visit mobility modeling to incorporate accurate behavioral models in the future. Our analysis is the first to provide insights leading to that direction.

## 2. Related Work

A new paradigm of protocols that relies on the human behavioral patterns has gained recent attention in DTN-related research. In these studies, researchers attempted to use social aspects of human mobility to derive new services and protocols. The study in [8] has used social network metrics for information flow and behavioral routing in DTNs and shown that using nodes with high betweenness as forwarders achieves best overall delivery performance. A similar study in [6, 20, 21] has used various centrality measures and proposed an array of forwarding algorithms for information dissemination in opportunistic networks. A new behavior-oriented service, called profile-cast, that relies on tight user-network integration was introduced in [15]. Profile-cast provides a systematic framework to utilize implicit relationships discovered among mobile users for interest-based message efficient forwarding and delivery in DTNs. Participatory sensing [35, 39, 43] provides a service for crowd sourcing using recruiting campaigns using mobile user profiles [16]. All these works rely on and utilize similarity of mobile user profiles.

Modeling and simulating profile similarity accurately in networked mobile societies is thus imperative to the design and evaluation of these classes of networking protocols and services. The aim of this paper is to identify the underlying similarity structure that govern mobile societies and provide measures of its existence in common mobility models both quantitative and qualitatively.

There is a large body of work on mobility modeling and trace analysis, but we shall only discuss briefly (for lack of space) some of the main approaches. Generally, mobility models are either synthetic or trace-driven. The most common models include random mobility generators, including random direction, random waypoint and random walk. In addition, several other models attempt to capture spatial correlation (e.g., group mobility), temporal correlation (e.g., freeway) and geographic restrictions (e.g., pathway, manhattan) [1]. These models are useful for simple initial evaluations, but not based on real traces and have been shown to lack important characteristics of user location preferences and periodicity [19]. Some community models [12, 34] represent social connections based on social network theory, but do not capture mobility related preferences of similarity. Trace-based mobility models, by contrast, derive their parameters from analysis of measurements and can reproduce user preferences [19]. There has been no prior evaluation, however, of similarity and clustering in mobility models. Recently, several studies have attempted to analyze realistic behavioral through processing of WLAN traces. In [26, 47] individual behavior (preferences) of users was studied, while in [17] nodal encounters have been analyzed. In [16] users preferences are captured through an association matrix and analyzed. We utilize a similar representation here. This study, however, is the first to compare trace-based similarity clustering to model-based similarity clustering and show the need to re-visit mobility modeling to capture similarity characteristics.

## 3. Similarity

The congregation of mobile agents with similar characteristic patterns naturally develop mobile societies in wireless networks [10, 22, 48]. Upon reflection it should come as no surprise that these characteristics in particular also have a big impact on the overall behavior of the system [7, 21, 32, 33]. Researchers have long been working to infer these characteristics and ways to measure them. One major observation is that people demonstrate periodic reappearances at certain locations [9, 16, 24], which in turn breeds connection among similar instances [31]. Thus, people with similar behavioral principle tie together. This brings an important aspect where, user-location coupling can be used to identify similarity patterns in mobile users. So, for the purpose of our study, to quantify similarity characteristics among mobile agents, we use their spatio-temporal preferences and preferential attachment to locations and the frequency and duration of visiting these locations.

It is important to study similarity in DTN to develop behavioral space for efficient message dissemination [15] and design behavior-aware trust advisors among others [28]. For efficient networking, it can help to quantify traffic patterns and develop new protocols and application to target social networking. Analysis of similarity can be used to evaluate the network transitivity, which helps to analyze macro-mobility, evolutionary characteristics and emergent properties.

In this section, we introduce association matrix that captures spatio-temporal preferences and a statistical technique that use it to measures similarity among mobile users.

### 3.1 Capturing Spatio-Temporal Preferences

We use longitudinal wireless activity session to build mobile user's spatio-temporal profile. An anonymous sample is shown in Table 1. Each entry of this measurement trace has the location of association and session time information for that user. The location association coupled with time dimension provides a good estimate of user online mobile activity and its physical proximity with respect to other online users [17, 26]. We devise a scalable representation of this information in form of an *association matrix* as shown in **Figure 1**. Each individual column corresponds to a unique location in the trace. Each row is an *n*-element association vector, where each entry in the vector represents the fraction of online time the mobile user spent at that location, during a certain time period (which can be flexibly chosen, such as an hour, a day, etc.). Thus for *n* distinct locations and *t* time periods, we generate a *t-by-n* size association matrix.

Table 1: Anonymized Sample of Mobile user WLAN session

| Node Mac ID | Location | Start Time | End Time |
|---|---|---|---|
| aa:bb:cc:dd:ee:ff | Loc-1 | 64400343 | 66404567 |
| aa:bb:cc:dd:ee:ff | Loc-2 | 85895623 | 86895742 |
| aa:bb:cc:dd:ee:ff | Loc-3 | 87444343 | 89404567 |
| aa:bb:cc:dd:ee:ff | Loc-4 | 98846767 | 99878766 |

*Representation Flexibility*: The representation of spatio-temporal preferences in form of an association matrix can be changed to use each column for a building (where a collection of access points represent a building) and the time granularity can be changed to represent hourly, weekly or monthly behavior. For the purpose of our study, each row represents a day in the trace and column represents an individual access point.

### 3.2 Characterizing Association Patterns

For a succinct measure of mobile user behavior, we capture the dominant behavioral patterns by using Singular Value Decomposition (SVD) [14] of the association matrix. SVD has several advantages: 1) It helps to convert high dimensional and high variable data set to lower dimensional space there by exposing the internal structure of the original data more clearly. 2) It is robust to noisy data and outliers. 3) It can easily be programmed for handheld devices, which is our other on-going work.

The Singular Value Decomposition of a given matrix *A* can be represented as a product of three matrices: an orthogonal matrix U, a diagonal matrix S, and the transpose of an orthogonal matrix V. It is written as:

$$A = U \cdot S \cdot V^T$$

Where, $U^T \cdot U = I = V^T V$, U is t-by-t matrix whose columns are orthonormal eigenvectors of $AA^T$, S is a t-by-n matrix with r non-zero entries on its main diagonal containing the square roots of eigen values of matrix A in descending order of magnitude and $V^T$ is a n-by-n matrix whose columns are the orthonormal eigenvectors of $A^T A$. Thus the eigen behavior vectors of $V = \{v_1, v_2, v_3, ... v_n\}$ summarize the important trends in the original matrix A. The singular values of $S = \{s_1, s_2, s_3, ..., s_r\}$ ordered by their magnitude $\{s_1 \geq s_2 \geq ... \geq s_r\}$. The percentage of power captured by each eigen vector of the matrix A is calculated by

$$w_i = \sum_{i=1}^{k} s_i^2 \Big/ \sum_{i=1}^{rank(A)} s_i^2$$

It has been shown that [16] SVD achieves great data reduction on the original association matrix and 90% or more power for most of the users is captured by five components of the association vectors. By this result, we infer that user's few top location-visiting preferences are more dominant than the remaining ones.

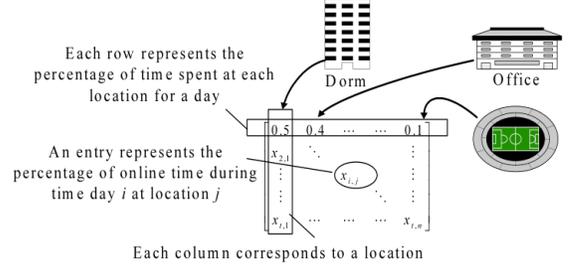

(a)

$$A = \begin{bmatrix} 0 & 0 & 0.5 & 0.3 & 0.2 \\ 0.2 & 0.3 & 0.4 & 0.1 & 0 \\ 0 & 0 & 0 & 0.6 & 0.4 \\ 0.2 & 0.3 & 0.1 & 0.2 & 0.2 \\ 0.1 & 0 & 0 & 0 & 0.9 \end{bmatrix}$$

(b)

Figure 1: (a) A prototype of *Association Matrix*. The columns represent locations (access point, building, etc) and rows represent time granularity (days, weeks, etc.). (b) A computed matrix *A* with 5 locations and time periods. Each entry represent the percentage online time spent at corresponding location column.

### 3.3 Calculating Similarity

We use the eigen vectors of association matrix *A* to quantitatively measure the similarity between behavioral profiles of mobile user pairs. For a pair of users, with respective eigen-vectors as $X = \{x_1, x_2, x_3, ... x_{r_x}\}$ and $Y = \{y_1, y_2, y_3, ... y_{r_y}\}$, the behavior similarity can be calculated by the weighted sum of pair wise inner product of their eigen vectors as

$$Sim(X, Y) = \sum_{i=1}^{rank(X)} \sum_{j=1}^{rank(Y)} w_{x_i} w_{y_j} |x_i \cdot y_j|$$

*Sim(X,Y)* is quantitative measure index that shows the closeness of two users in spatio-temporal dimension. The value of similarity lies between $0 \leq Sim(X, Y) \leq 1$. A higher value is derived from users with similar association patterns. In this study, we are the first one to investigate the distribution of such a similarity metric among user pairs based on realistic data sets.

### 4. Trace-Based Similarity Analysis

In this section, we obtain the distribution of Similarity among mobile users and apply a divisive quality function – *modularity* [36] to discover mobile societies in wireless networks. We expect to see a natural division of mobile nodes into densely connected clusters. Each of these clusters consists of users with similar spatio-temporal preferences. For the purpose of this study, we examine very large real world wireless data measurements of four university campuses collected for a period of several months with thousands of users.

### 4.1 Dataset and Trace Analysis

WLAN dataset from four university campuses are considered as shown in Table 2. We collect these datasets from the publicly available MobiLib[38] and Crawdad[27] repositories. Table 2 provides the detail of these WLAN measurements. We chose university campuses because they are extensive, have high density

**Table 2: Details of Wireless Measurements**

| Campus | # Users | Duration |
|---|---|---|
| Dartmouth | 1500 | Fall 2007 |
| MIT | 1366 | Fall 2006 |
| Univ. of Florida | 3000 | Fall 2008 |
| USC | 3000 | Fall 2007 |

## 4.2 Similarity Analysis

The distribution histogram of similarity scores for the campus datasets is shown in **Figure 2**. The figure shows number of user pairs as a function of similarity score that quantify the behavioral similarity between mobile users. We observe that: 1) mobile societies compose of users with mixed behavioral similarities, 2) For all four time periods there is a consistency and stability in the similarity score among mobile user pairs. The low similarity scores (0 - 0.1) in **Figure 2** indicate a substantial portion of users is spatio-temporally very dissimilar. On the other hand, similarity scores of (0.9 – 1.0) suggest a statistically significant likelihood of high-density ties creating tightly knit groups. The variation in the middle shows partially similar and partially dissimilar user pairs. This is significant and provides an insight into the existence of mobile societies in the network with quite similar location visiting preferences. Overall, the curves show an assortative mixing of user pairs for all possible similarity scores. **Figure 3** gives a normalized log plot to compare data sets from of active users and include location information. Also, these datasets have been used in previous studies of mobility modeling [19, 21, 24, 48]. We perform Systematic Random Sampling [42] on the datasets to get an unbiased subset of mobile users from the population. Table 2 specifies the sampling frame that we use for this study. In the second step, we extract relevant statistics of mobile user spatio-temporal patterns. In the third step, for each mobile user we obtain normalized association matrix as shown in Figure 1 with time granularity of one day. On this matrix we apply SVD to extract the dominant trends. Finally, we compute the cosine similarity of all user pairs. We perform this process iteratively for four different time intervals: 1 week, 2 weeks, 3 weeks and 4 weeks.

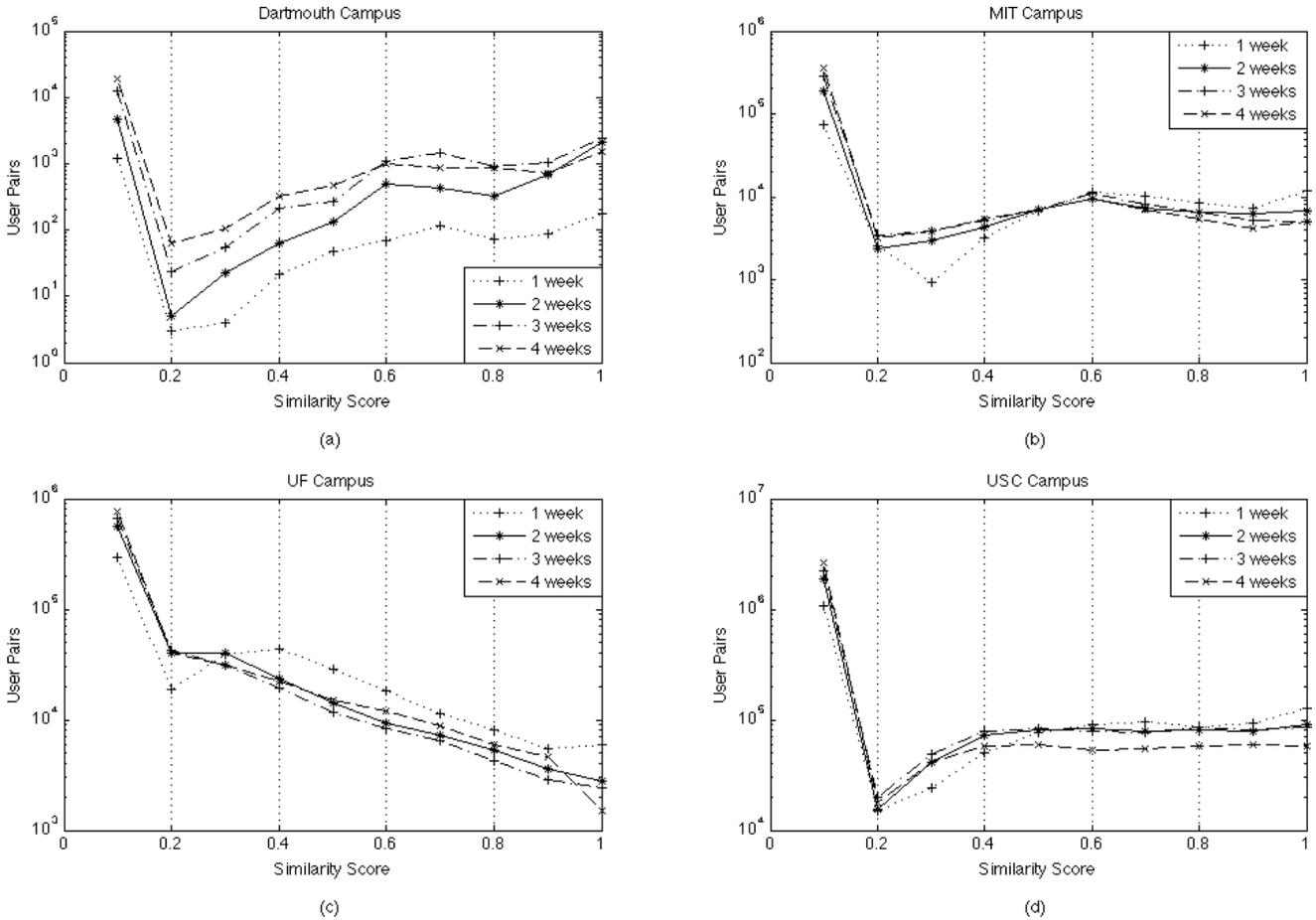

Figure 2: Similarity distribution histogram among user pairs is shown. All the four time intervals show near consistent user pair counts for a particular similarity score. Lowest similarity score (0.0 - 0.1) shows that users have very different spatio-temporal preferences. A fraction of the user pairs are also very similar with (0.9 - 1.0) similarity score.

different campuses, and shows that similarity exists evenly across all the traces. Next, we briefly explain modularity and use a divisive algorithm to discover mobile societies in the traces.

Table 3: Network Analysis of Datasets on three different metrics

| Dataset | Clustering Coefficient | | Average Path Length | | Modularity | |
|---|---|---|---|---|---|---|
| | *Ori* | *Rand* | *Ori* | *Rand* | *Ori* | *Rand* |
| Dartmouth | 0.89 | 0.05 | 0.10 | 2.47 | 0.63 | 0.2 |
| MIT | 0.92 | 0.05 | 0.40 | 2.12 | 0.79 | 0.14 |
| UF | 0.78 | 0.051 | 0.30 | 2.605 | 0.67 | 0.24 |
| USC | 0.91 | .05 | 0.19 | 2.0 | 0.46 | 0.11 |
| *Ori = Original Dataset Graph* | | | *Rand = Random Graph* | | | |

### 4.2.1 Modularity

To understand the underlying structure of mobile societies (or communities), the similarity distribution is not sufficient. Therefore, we use a robust method to segregate user pairs that have high similarity score into tightly knit groups. To detect such communities in a graph like structure, a centrality-index-driven method [13] is utilized. This measure to detect communities circumvents the traditional clustering notion to identify most central edges. Instead, a divisive algorithm is applied based on identifying least central edges, which connect most communities (via *edge betweenness*). First, the *betweenness score* of edges are calculated as the number of shortest paths between pair of vertices that run through it. Understandably, tightly knit communities are loosely connected by only few intergroup edges and hence shortest paths traverse these edges repeatedly, thereby increasing their respective *betweenness* score. If such edges are removed, according to a threshold, what we get are the groups of tightly knitted vertices known as communities. To identify a reasonable threshold value, *modularity* is used. Modularity is the difference of edges falling within communities and the expected number in an equivalent network with randomly-placed edges [13, 36, 37].

### 4.2.2 Detection of Mobile Societies

Human networks are known to exhibit a multitude of emergent properties that characterize the collective dynamics of a complex system [23, 45, 49]. Their ability to naturally evolve into groups and communities is the reason they show non-trivial clustering. Here, we consider the spatio-temporal preferences and cosine similarity of mobile users as a relative index to generate emergent structures, which we call *mobile societies*. The network transitivity structures of mobile nodes for various campus datasets are shown in **Figure 4**. We use mutual similarity score of mobile nodes to produce a connected graph and applied random iterations of modularity [3] and betweenness algorithm to infer the mobile societies. A set of visibly segregated clusters validates their detection and presence in mobile networks.

#### 4.2.2.1 Modularity Analysis for Mobile Societies

Statistically, modularity greater than 0.4 is considered meaningful in detecting community structure. For our dataset, we also find high modularity index as compared to an equivalent random graph. The comparison is shown in Table 3. Henceforth, the heterogeneity in dataset has tightly knitted Mobile Societies. This analysis further helped us to investigate the possibility of existence of different clusters of users based on their proximity in similarity score values.

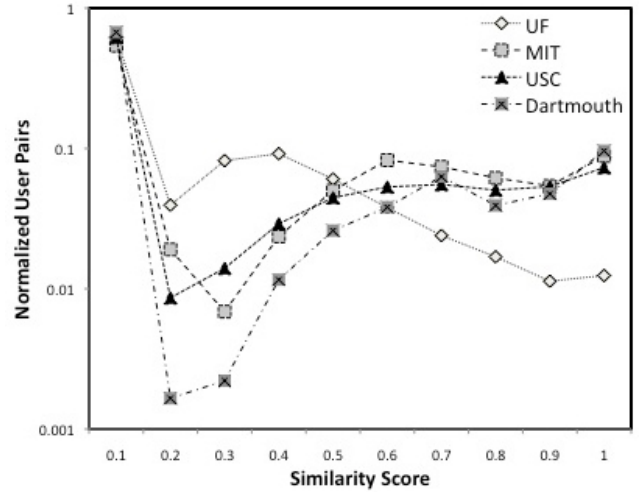

Figure 3: Log Normalized Similarity distribution of all four data sets is shown.

#### 4.2.2.2 Network Analysis for Mobile Societies

We compute the average clustering coefficient and the mean-shortest path length of these clusters. We compare the results with a random graph of the same size to understand the variation and capacity to depict small world characteristics. Table 3, delineates network properties and average modularity that provide details of the structure of mobile societies against same size random graph. The comparative values in the table clearly show that mobile societies can exhibit small world characteristics. However, we leave such small world study for future work.

Based on the above analysis, we find that similarity not only exists among mobile users, but its distributions seem to be stable for different time periods. Furthermore, this trend is consistent in all four traces, which highlights similarity clustering as an important characteristic to capture using mobility models.

## 5. Similarity in Models - The Missing Link

In this section, we evaluate existing mobility models and contrast their output against real trace results. Trace based mobility models [4, 6, 11, 18, 25, 29, 30, 40] are a close approximation of realistic human movements and their non-homogenous behavior. They focus on vital mobility properties like node's on/off behavior, connectivity patterns, spatial preferences under geographical restrictions, contact duration, inter-meeting and pause time, etc. We consider two mobility models, the random direction model (a widely used "classic" mobility model) and Time Variant Community Model [19] (due to its capability to capture spatio-temporal mobility properties). In the ensuing text, we briefly describe the TVC model and use it to generate realistic movements. Finally, we compare its result against the similarity characteristic found in real measurements.

### 5.1 TVC Model

The TVC model [19] is proposed to capture two prominent features in wireless network user mobility observed in real traces (1) skewed location visiting preferences and (2) periodical re-appearance at the same location. TVC model introduces multiple "preferred geographical locations", or the "communities", to which the mobile nodes visit often, in order to capture the fact that most mobile users spend significant portion of time at a few locations. Further, the TVC model also introduces a structure in time (the "time periods") that allows setting up different mobility preferences for users in a periodic fashion. TVC model is the best we have found in literature in terms of the capacity to closely reproduce realistic user behavior and the flexibility to be fine-tuned for different environments. Secondly, TVC also encompass properties depicted by [41, 44], which implicitly helps us to evaluate similarity on these models also. Thus, we adopt the TVC model in this study and evaluate its capacity to capture the social structure we observed in section 4.

curves for both campuses. We clearly observe a discrepancy between the curves from actual traces and the two mobility models (TVC and random direction). In addtion, dendrograms in **Figure 6** shows the result of hierarchical clustering based on user's mutual similarity scores. Here, in real traces we find clusters at different similarity scores. In Figure 6(a), the average distance of 2.0 has close to 18 small clusters and Figure 6(c) shows 16 small clusters of mobile users. However, corresponding TVC

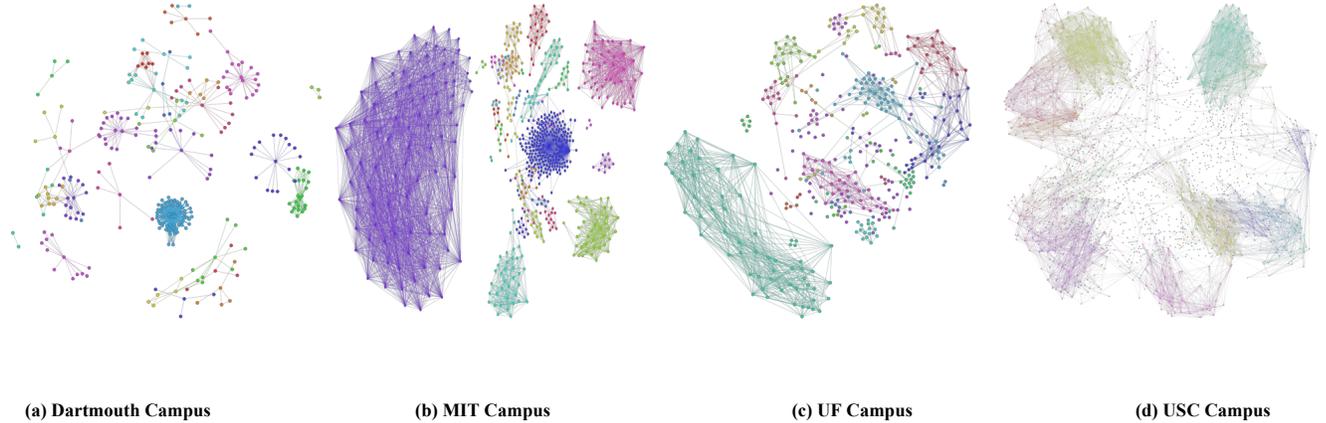

(a) Dartmouth Campus  (b) MIT Campus  (c) UF Campus  (d) USC Campus

Figure 4: Shown are the structural and spatio-temporal dynamics of Mobile Societies as function of weighted cosine similarity score, produced from highly positive modularity values. *Note: this figure is best viewed in color.*

## 5.2 TVC Model Evaluation

We setup the TVC model for two university campuses (MIT and USC) to statistically evaluate the similarity metric established previously. Our goal is two folds:

1. As proposed by the TVC model, we seek to maintain the skewed location visiting preferences and time dependent mobility behavior of users.
2. To analyze whether TVC model successfully captures similarity among mobile users and quantitatively simulate the distribution that we have seen in the real measurements.

### 5.2.1 Construction of TVC Model for Campuses

Initially, we determine the number of communities that nodes should periodically visit. We determine that top 2-3 communities capture most skewed location visiting preferences. Then we employ a weekly time schedule to capture the periodic re-visits to these major communities. To keep fair comparison against the real measurements, we configure the TVC model with same number of mobile nodes and generating measurements equivalent to one month time period with one-day granularity. Finally, for WLAN measurement we assume mobile users are stationary while being online [19].

### 5.2.2 Similarity Evaluation

TVC model accurately demonstrates location visiting preferences and periodic reappearances for both campuses [46]. Surprisingly, it is unable to accurately capture the richness in similarity distribution on spatio-temporal basis. For all values of similarity score except 0.9, TVC and Random Direction model yields no user pairs. Figure 5 shows similarity distribution CDF dendrograms in Figure 6(b) and 6(d) show only one cluster of mobile users at a distance of 2.0. A possible explanation is that the community assignment in TVC model creates a homogeneous user population where all nodes are assigned the same communities. While it captures the location visiting and periodic preferences, it fails to differentiate among mobile nodes with different behaviors. What is missing here is a mechanism to assign different locations as the communities to different nodes, in a way that reproduces the social structure (clusters) observed in the traces.

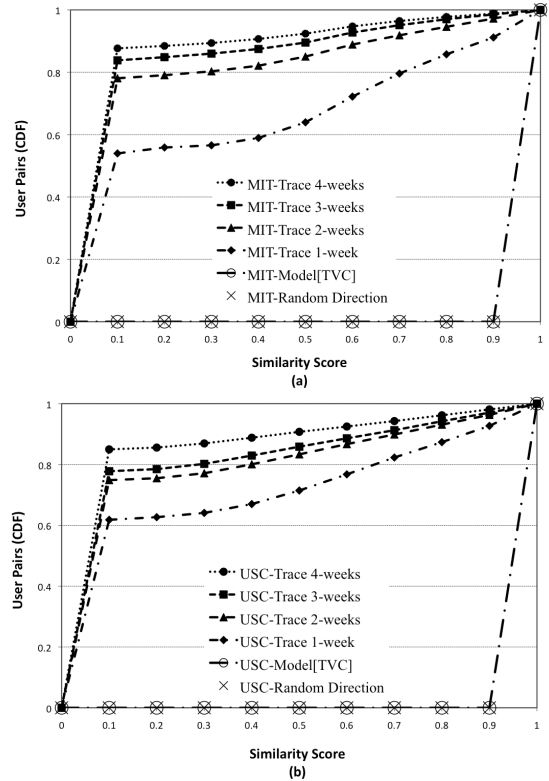

Figure 5: Cumulative distribution function of distances for the similarity score of mobile users. Real trace curves show a conformance with user pairs for different values of similarity score, while TVC Model has all users pairs in the 0.9 score range.

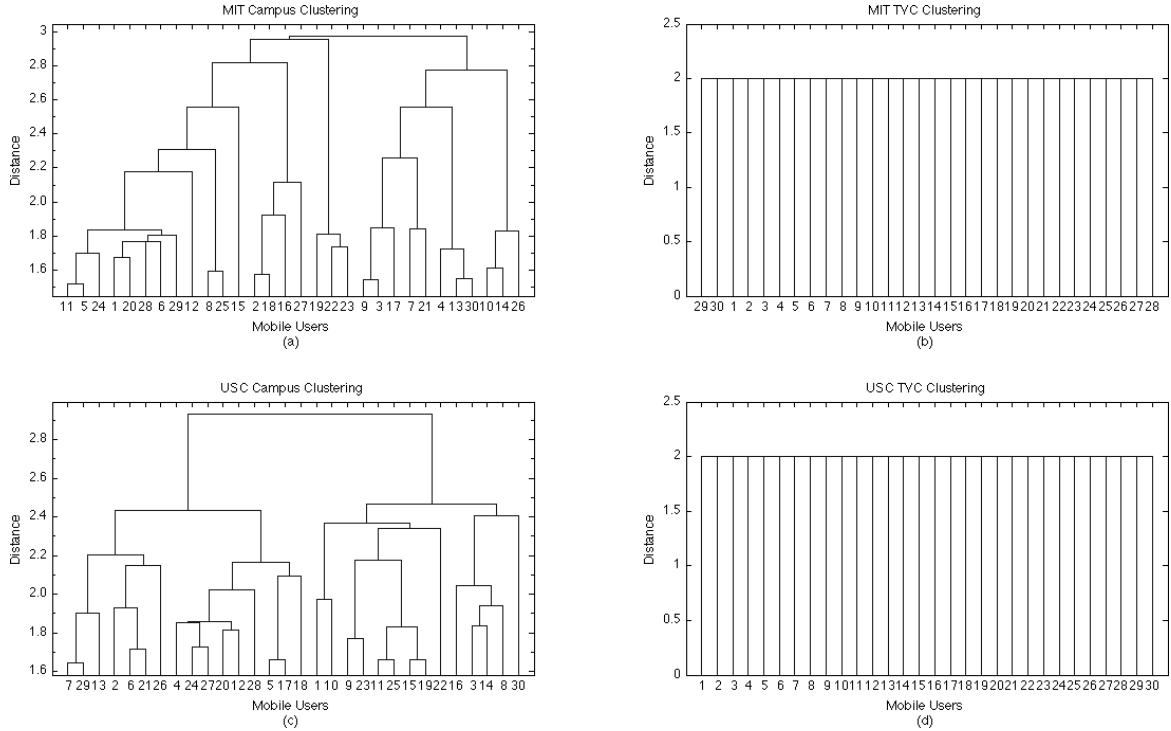

Figure 6: Dendrograms giving visual representation of two-dimensional hierarchical clustering for real and TVC model generated traces for USC and MIT campus mobile users. Real traces (Figure a & c) show an incremental built-up of component based on the similarity score strength between mobile user. TVC Model (Figure b & d), output only one cluster containing all mobile users. Invariably, TVC treats all mobile users to have same preferences.

Results in this section show that although TVC model is able to capture location visiting preferences and periodic reappearances, it does not capture the similarity metric distribution and the clusters with different behaviors in the traces. Random direction model also fails on this front in a similar way. This study realizes us that current mobility models are not fully equipped to handle behavioral metrics and community behavior of users that form mobile societies. It compels us to revisit mobility modeling in the attempt to capture both individual and community behavior of mobile users, which is part of our on-going study.

## 6. Conclusion and Future Work

Several novel behavior-aware protocols and services are being designed for DTNs, using similarity as a corner stone for their architecture. In this paper, we analyze the spatio-temporal behavioral similarity profiles among mobile users. We define mobility profiles based on users association matrices, then use a SVD-based-weighted-cosine similarity index to quantitatively compare these mobility profiles. Analysis of extensive WLAN traces from four major campuses reveals rich similarity distribution histograms suggesting a clustered underlying structure. Application of modularity based clustering validated and further quantified the clustered behavior in mobile societies. Similarity graphs exhibit an average modularity of 0.64, and clustering coefficient of 0.86, which indicates potential for further small world analysis. Finally, we compared similarity characteristics of the traces to those from existing common and community based mobility models to capture similarity. Surprisingly, existing models are found to generate a homogeneous community with one cluster and thus deviate dramatically from realistic similarity structures. This indicates a serious flaw in existing models and their lack of support for emerging behavior-aware protocols, and provides a compelling motivation to re-visit mobility modeling. We believe this is vital for the evaluation and design of next-generation behavior-aware protocols likely to be used in DTN and other ad hoc networks.

In the future, we plan to further investigate similarity modeling and its effect on protocol efficiency and routing decisions in mobile networks. We also plan to perform stability, evolution and sensitivity analysis of similarity across different time and space granularities. Furthermore, we shall introduce a new mobility model with a systematic procedure to capture realistic similarity structures in mobile societies. The key insight, based on this study, is to assign communal probabilities to nodes using explicit parameters to capture inter-dependencies between behavioral similarity clusters.